\documentclass{article}
\usepackage[english]{babel}
\usepackage[latin1]{inputenc}

\usepackage{spconf,amsmath,graphicx}
\usepackage{amsthm}
\usepackage{amssymb}
\usepackage{slashbox}
\usepackage{bbm}
\usepackage{array}
\newcommand\Mark[1]{\textsuperscript#1}

\title{DISTRIBUTED DIFFUSION-BASED LMS FOR NODE-SPECIFIC PARAMETER ESTIMATION OVER ADAPTIVE NETWORKS}
\name{Nikola Bogdanovi\'{c}\Mark{1}, Jorge Plata-Chaves\Mark{2}, Kostas Berberidis\Mark{1} \thanks{The  work was partially supported  by the European HANDiCAMS project (Grant No. 323944) under the Future and Emerging Technologies (FET) programme within the Seventh Framework Programme for Research of the European Commission, in part by the EU
and national funds via the National Strategic Reference Framework (NSRF)-
Research Funding Program: Thales (ENDECON) and in part by the University of Patras.} }

 \address{\Mark{1}Department of Computer Engineering and Informatics,  University of Patras \\ \& C.T.I RU-8, 26500, Rio - Patra, Greece\\
\Mark{2}Department of Electrical Engineering-ESAT, STADIUS, KU Leuven, B-3001 Leuven, Belgium\\
 E-mails: \{bogdanovic, berberid\}@ceid.upatras.gr; jplata@esat.kuleuven.be}

\begin{document}
\ninept
\maketitle
\begin{abstract}

A distributed adaptive algorithm is proposed to solve a node-specific parameter estimation problem where nodes are interested in estimating parameters of local interest and parameters of global interest to the whole network. To address the different node-specific parameter estimation problems, this novel algorithm relies on a diffusion-based implementation of different Least Mean Squares (LMS) algorithms, each associated with the estimation of a specific set of local or global parameters. Although all the different LMS algorithms are coupled, the diffusion-based implementation of each LMS algorithm is exclusively undertaken by the nodes of the network interested in a specific set of local or global parameters. To illustrate the effectiveness of the proposed technique we provide simulation results in the context of cooperative spectrum sensing in cognitive radio networks. 

\end{abstract}

\section{Introduction}
\label{sec:sec1}

Two major groups of energy aware and low-complex distributed strategies for estimation over networks have been studied in the literature, i.e., consensus strategies and the algorithms based on incremental or diffusion mode of cooperation.
In some initial works, for instance~\cite{olfati2004}, 
the implementation of the consensus strategy is done in two stages.  
Unfortunately, this kind of implementation is not suitable for real time estimation as required in time-varying environments. Subsequently, motivated by the procedure obtained in 
\cite{bertsekasparallel}, alternative implementations of the consensus strategy were presented in the literature (e.g., ~\cite{schizas2009distributed}
-\cite{dimakis2010gossip}) which force  agreement among the cooperating nodes in a single time-scale. The second group, which is in the focus of this paper, consists of a single time-scale distributed estimation algorithms that are based on 
distributing a specific stochastic gradient method under an incremental or a diffusion mode of cooperation.  In the incremental mode (e.g., 
\cite{lopes2007incremental}-\cite{li2010distributed}), 
each node communicates with only one neighbor, and consequently the data are processed in a cyclic manner throughout the network. 
Better reliability can be achieved at the expense of increased energy consumption in the so-called diffusion mode considered, for instance, in 
\cite{cattivelli2010diffusion}\nocite{chouvardas2011}-\cite{sayed2013_magazine}. Under this strategy, each node can communicate with a subset of neighboring nodes.

Although there are many published techniques addressing different distributed estimation problems, only very  few papers consider node-specific settings where the nodes have overlapped but different estimation interests.
 In the signal estimation case, for networks with a fully connected and tree topology, Bertrand \emph{et al}. proposed distributed algorithms that allow to estimate node-specific desired signals sharing a common latent signal subspace~(\cite{bertrand2010distributed}-\cite{bertrand2010distributed_smlt}).
Regarding the parameter estimation case,
there are also a few recent works addressing problems which can be considered as Node-Specific Parameter Estimation (NSPE) problems.  The consensus approach presented in~\cite{kekatos2012distributed} is based on optimization techniques that force different nodes to reach an agreement when estimating parameters of common interest. In the case of schemes based on a distributed implementation of adaptive filtering techniques, the literature is less extensive. In one of these works~\cite{chen2012distributed}, the authors use diffusion adaption and scalarization techniques to solve the multi-objective cost function that appears in a NSPE problem and obtain a Pareto-optimal solution.
A diffusion strategy with an adaptive combination rule proposed in~\cite{zhao2012clustering} is suitable for clustering nodes in a network that are interested in different objectives. Consequently, 
it actually limits cooperation only to the nodes having exactly the same objectives. In~\cite{abdolee2012diffusion}, the authors assume a NSPE setting, however, the different parameters to be estimated using diffusion strategy are expressed through the same global parameter. 
In previous works~\cite{bogdanovic2013a} and~\cite{platachaves2013a}, we formulated a novel NSPE problem where all nodes are interested in estimating simultaneously some parameters of local interest as well as some parameters of global interest. We solved it by employing incremental-based strategies for Least Mean Squares (LMS) and Recursive Least Squares (RLS) algorithms. Motivated by the well-known robustness and learning abilities of the diffusion-based solutions, in this work we present a LMS strategy to solve the aforementioned NSPE problem under two different versions of the diffusion mode of cooperation, Combine-then-Adapt and Adapt-then-Combine. 
Finally, we verify the effectiveness of both techniques through an illustrative application for power spectrum sensing in cognitive radio.

The following notation is used throughout the paper. We use boldface letters for random variables and normal fonts for deterministic quantities. Capital letters refer to matrices and small letters refer to both vectors and scalars. The notation $(\cdot)^H$ and $E\{ \cdot \}$ stand for the Hermitian transposition and the expectation operator, respectively. 
Moreover, $R_{\mathbf{A}} = E \{\mathbf{A}^H \mathbf{A}\}$, $R_{\mathbf{A},\mathbf{B}} = E \{\mathbf{A}^H \mathbf{B}\}$ and $r_{\mathbf{A},\mathbf{b}} = E \{\mathbf{A}^H \mathbf{b}\}$ for any random matrices $\mathbf{A}$, $\mathbf{B}$ and any random vector $\mathbf{b}$. Finally, $\Vert \cdot \Vert$ denotes the Euclidean norm and $0_{L \times M}$ represents a $L \times M$ zero matrix.

\section{Problem statement}
\label{sec:sec2}

\begin{figure}[t]
\begin{minipage}[b]{1.0\linewidth}
  \centering
  \centerline{\includegraphics[width=5.5cm]{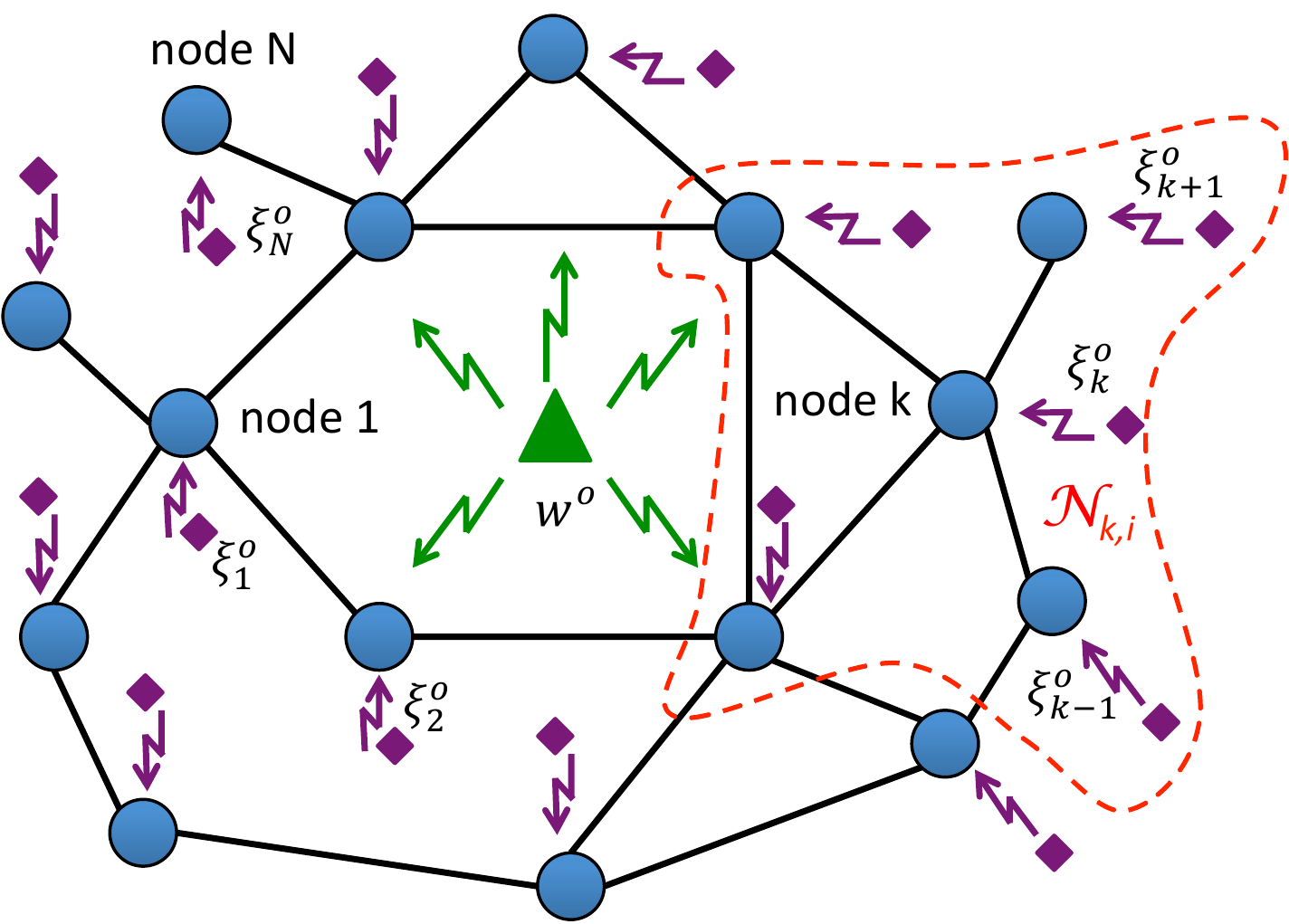}}
  \vspace{-0.15cm}
\end{minipage}
\caption{ A network of $N$ nodes with node-specific parameter estimation interests.} 
\label{fig:fig1}
 \vspace{-0.25cm}
\end{figure}

Let us consider a connected network consisting of $N$ nodes (~Fig.\ref{fig:fig1}). 
Hence, allowing each node to communicate with its neighbors, at each time instant there is always a path between any two pairs of nodes of the network. As shown in Fig.~\ref{fig:fig1}, the neighborhood of a node $k$ at a specific time instant $i$, $\mathcal{N}_{k,i}$, consists of the nodes linked to it, including node $k$ itself. 

At discrete time $i$, each node $k$ has access to data $ \{d_{k,i}, U_{k,i} \}$, corresponding to time realizations of zero-mean random processes $\{\mathbf{d}_{k,i}, \mathbf{U}_{k,i} \}$, with dimensions $L_k \times 1$ and $L_k \times M_k$, respectively. %
%
%
These data are related to events that take place in the network area 
through the subsequent model
\begin{gather}\label{eps:eq1}
\begin{split}
\mathbf{d}_{k,i} &= \mathbf{U}_{k,i} w_k^o + \mathbf{v}_{k,i} 
\end{split}
\end{gather}
where, for each time instant $i$, $w_k^o$ equals the $M_k \times 1$ vector 
that gathers all parameters of interest for node $k$ and $\mathbf{v}_{k,i}$ denotes measurement and/or model noise with zero mean and covariance matrix $R_{v_k,i}$ of dimensions $L_k \times L_k$.

Given the previous observation model and the data set $\{d_{k,i}, U_{k,i}\}$, 
the objective is to find the set of linear node-specific estimators $\{w_k\}_{k=1}^{N}$ that minimize the following global cost function
\begin{gather}\label{eps:eq2}
\begin{split}
J_{\textrm{glob}}(\{w_k\}_{k=1}^{N}) &=\sum_{k=1}^N E \left \{ \Vert \mathbf{d}_{k,i} - \mathbf{U}_{k,i} w_k \Vert^2 \right \}.
\end{split}
\end{gather}
In most of the existing papers,  e.g.,~\cite{lopes2007incremental}\nocite{li2010distributed}\nocite{cattivelli2010diffusion}\nocite{chouvardas2011}-\cite{sayed2013_magazine}, the derived adaptation strategies minimize~\eqref{eps:eq2} when $w_k^o = w^o$ for all $k \in \{1,2,\ldots,N\}$. 
In this work, we consider the novel node-specific parameter setting addressed in~\cite{bogdanovic2013a} and~\cite{platachaves2013a}, which goes one step further by considering a more general scenario where the parameters of interest can differ from one node to another. 
As shown in Fig.~\ref{fig:fig1}, each vector $\{w_k^o\}_{k=1}^N$ might consist of parameters of global interest to the whole network and parameters of local interest for node $k$. In particular, the global parameters might be related to a phenomenon making an impact on all the nodes, 
while the parameters of local interest may reflect an influence of some phenomena that are only present over the area monitored by one node of the network. Therefore, we can rewrite the observation model in~\eqref{eps:eq1}, for each node $k$, as
\begin{gather}\label{eps:eq3}
\begin{split}
\mathbf{d}_{k,i} 
&= \mathbf{U}_{k_g,i} w^o +  
\mathbf{U}_{k_l,i} \xi_k^o  + \mathbf{v}_{k}^{(i)} 
\end{split}
\end{gather}
where  sub-vectors $w^o$ ($M_{g} \times 1$) 
and $\xi_k^o$ ($M_{k_l} \times 1$) gather the parameters of global 
and local interest, respectively. Furthermore,  $\mathbf{U}_{k_g,i}$ 
and $\mathbf{U}_{k_l,i}$ are matrices of dimensions $L_k \times M_{g}$ 
and $L_k \times M_{k_l}$ that consist of the columns $\mathbf{U}_{k,i}$ associated with $w^o$ 
and $\xi_k^o$, respectively. 
Thus, according to~\eqref{eps:eq2} and~\eqref{eps:eq3}, our NSPE problem can be cast as minimizing 
\begin{gather}\label{eps:eq4}
\begin{split}
 \sum_{k=1}^N E \left \{ \Vert \mathbf{d}_{k,i} -  \mathbf{U}_{k_g,i} w - 
\mathbf{U}_{k_l,i} \xi_k \Vert^2 \right \}.
\end{split}
\end{gather}
with respect to $w$ and $\{\xi_k\}_{k=1}^{N}$. In the following section, we write the centralized solution to~\eqref{eps:eq4} and later, approximate it in a distributed manner via diffusion-based approach.

\section{Diffusion-based LMS for the NSPE problem}
\label{sec:sec3}

For simplicity and without losing generality, let us assume that $M_k=M$, $M_{k_l}=M_l$ and $L_k=L$ for all $k \in \{1,2,\ldots,N\}$. From~\cite{bogdanovic2013a}, we know that the our NSPE problem can be cast as
\begin{gather}\label{eps:eq5}
\begin{split}
\widehat{\tilde{w}} 
&=\underset{ \tilde{w}  }{\mathrm{argmin}} \left \{ \sum_{k=1}^N E \left \{ \Vert \mathbf{d}_{k,i} -  \mathbf{\widetilde{U}}_{k,i} \tilde{w} \Vert^2 \right \} \right \}
\end{split}
\end{gather}
where
\begin{gather}\label{eps:eq7}
\begin{split}
\tilde{w}&=\left [w^T \,
 \xi_1^T \, \,  \xi_2^T    \cdots \,  \xi_N^T \right ]^T \quad ( \, \widetilde{M} \times 1 \, )
\end{split}
\end{gather}
and
\begin{gather}\label{eps:eq6}
\begin{split}
\mathbf{\widetilde{U}}_{k,i} = 
\begin{bmatrix}
\mathbf{U}_{k_g,i} 
& 0_{L \times M_{a}} & \mathbf{U}_{k_l,i} & 0_{L \times M_{b}}
\end{bmatrix}
\end{split}
\end{gather}
where $M_{a} =  (k-1) M_l$, $M_{b}=(N-k) M_l$ and $\widetilde{M}=M_g +N \cdot M_l$. Thus, the resulting solutions $\widehat{\tilde{w}}$ are given by the normal equations~\cite{sayed2011adaptive}, i.e.,
\begin{gather}\label{eps:eq9}
\begin{split}
\left( \sum_{k=1}^N R_{\widetilde{U}_{k,i}} \right ) \cdot \widehat{\tilde{w}} =  \sum_{k=1}^N  r_{\widetilde{U}_{k,i}d_{k,i} }.
\end{split}
\end{gather}

With the aim of improving energy efficiency, robustness and scalability of the centralized approach, it is highly desirable to design a distributed and adaptive scheme that allows each node to solve its NSPE problem. In case that $w^o_k=w^o$, diffusion strategies, e.g., Combine-then-Adapt (CTA) and Adapt-then-Combine (ATC), are known to well approximate the corresponding centralized solution by relying solely on information available at each node from its neighborhood \cite{cattivelli2010diffusion}. In this work, we extend these strategies so as to be applicable in the NSPE case.

To start with the derivation of the algorithm, let us define $\tilde{\psi}_k^{(i)}$ as the local estimate of $\tilde{w}^o$ at time instant $i$ and node $k$. Note that this local estimate is generally a noisy version of the optimal augmented vector $\tilde{w}^o$. 
By employing a diffusion mode of cooperation, each node $k$, at each time instant $i-1$, has access to the set of local estimates from its neighborhood, i.e., $\mathcal{N}_{k,i-1}$.
Hence, node $k$ can fuse its local estimate with the local estimates of its neighbors, at each time instant $i-1$, through a linear combiner as follows 
\begin{gather}\label{eps:eq12}
\begin{split}
\tilde{\phi}_{k}^{(i-1)}=\sum_{j \in \mathcal{N}_{k,i-1}} \widetilde{C}_{k,j} \, \tilde{\psi}_{j}^{(i-1)}
\end{split}
\end{gather}
where
\begin{gather}\label{eps:eq13}
\begin{split}
\widetilde{C}_{k,j}  = \mathrm{diag}\{ & c^{ w}_{k,j} I_{M_g}, 
 c_{k,j}^{\xi_{1}} I_{M_l}, \ldots,c_{k,j}^{\xi_{N}} I_{M_l} \}.
\end{split}
\end{gather} 
In~\eqref{eps:eq13}, $c^{ w}_{k,j}$ equals the weight coefficient used by node $k$ when combining the local estimate of the global vector $w^o$ from node $j$. Similarly, 
$c_{k,j}^{\xi_{m}}$ denotes the combination coefficients employed by node $k$ when fusing the local estimates of 
$\xi_{m}^{o}$, where $m \in \{1,2,\ldots,N\}$,  from node $j$ with its local estimates, respectively. 

To determine the combination coefficients at each node $k$, we can interpret~\eqref{eps:eq12} as a weighted least squares estimate of the augmented vector of parameters $\tilde{w}^{o}$ given its local estimate as well as the local estimates from the neighbor nodes~\cite{lopes2008diffusion}. This way, by collecting the local estimates of 
$\tilde{w}^o$ in the neighborhood of node $k$
\begin{gather}\label{eps:eq14}
\begin{split}
\tilde{\psi}_{\mathcal{N}_{k,i-1}} = \mathrm{col} \left \{ \{ \tilde{\psi}_{j}^{(i-1)} \}_{j \in \mathcal{N}_{k,i-1}} \right \}
\end{split}
\end{gather}
and defining 
\begin{gather}\label{eps:eq16}
\begin{split}
Q_{k,i-1} = \mathrm{col} \{I_{\widetilde{M}}, I_{\widetilde{M}}, \ldots, I_{\widetilde{M}} \} \quad \quad (n_{k,i-1}\cdot\widetilde{M} \times \widetilde{M})
\end{split}
\end{gather}
and $\widetilde{C}_k=\mathrm{diag} \{\widetilde{C}_{k,1}, \widetilde{C}_{k,2}, \ldots, \widetilde{C}_{k,n_{k,i-1}}\}$
with $n_{k,i-1}=|\mathcal{N}_{k,i-1}|$, we can formulate the subsequent local weighted least-squares problem
\begin{gather}\label{eps:eq15}
\begin{split}
\underset{ \tilde{\phi}_k  }{\mathrm{argmin}} \left \{ \Vert \tilde{\psi}_{\mathcal{N}_{k,i-1}} -  Q_{k,i-1} \tilde{\phi}_k \Vert^2_{\widetilde{C}_k} \right \},
\end{split}
\end{gather}
whose solution is given by
\begin{gather}\label{eps:eq18}
\begin{split}
\tilde{\phi}_k^{(i-1)} = \left [ Q^{T}_{k,i-1} \widetilde{C}_k Q_{k,i-1} \right ]^{-1} Q^{T}_{k,i-1} \widetilde{C}_k \tilde{\psi}_{\mathcal{N}_{k,i-1}}.
\end{split}
\end{gather}
More precisely, focusing on the different subvectors that form $\tilde{\phi}_k^{(i-1)}$, the solution provided in~\eqref{eps:eq18} can be rewritten as
\begin{gather}\label{eps:eq19}
\begin{split}
\phi_{k,w}^{(i-1)} = \sum_{j \in \mathcal{N}_{k,i-1} } \frac{c^{ w}_{k,j}}{\sum_{\ell \in \mathcal{N}_{k,i-1}} c_{k,\ell}^{w}} \psi_{j,w}^{(i-1)} 
\end{split}
\end{gather}
and
\begin{gather}\label{eps:eq21}
\begin{split}
\phi_{k,\xi_{m}}^{(i-1)} = \sum_{j \in \mathcal{N}_{k,i-1} } \frac{c_{k,j}^{\xi_{m}}}{\sum_{\ell \in \mathcal{N}_{k,i-1}} c_{k,\ell}^{\xi_{m}}} \psi_{j,\xi_{m}}^{(i-1)} 
\end{split}
\end{gather}
where, for $k,j,m \in \{1,2,\ldots,N\}$, $\phi_{k,w}^{(i-1)}$ 
and $\phi_{k,\xi_{m}}^{(i-1)}$ denote the subvectors of combiner $\tilde{\phi}_k^{(i-1)}$ associated with the local estimation of $w^o$ 
and $\xi_{m}$ at node $k$ and time instant $i-1$, respectively. Analogously, $\psi_{j,w}^{(i-1)}$ 
and $\psi_{j,\xi_{m}}^{(i-1)}$ denote the subvectors of local estimate $\tilde{\psi}_j^{(i-1)}$ associated with the local estimation of $w^o$ 
and $\xi_{m}^o$ at node $j$ and time instant $i-1$, respectively.

At this point,  after a suitable re-writing  of the combination coefficients that appear in~\eqref{eps:eq19} 
and~\eqref{eps:eq21}, we can verify that the combination coefficients in~\eqref{eps:eq12} and~\eqref{eps:eq13} have to satisfy
\begin{gather}\label{eps:eq22}
\begin{split}
c^{ w}_{k,j} = 0 \textrm{ if } j \notin \mathcal{N}_{k,i-1}; \quad \sum_{j \in \mathcal{N}_{k,i-1}} c^{ w}_{k,j} = 1
\end{split}
\end{gather}
and
\begin{gather}\label{eps:eq24}
\begin{split}
c_{k,j}^{\xi_{m}} = 0 \textrm{ if } j \notin \mathcal{N}_{k,i-1}; \quad \sum_{j \in \mathcal{N}_{k,i-1}} c_{k,j}^{\xi_{m}} = 1
\end{split}
\end{gather}
for $k,m \in \{1,2,\ldots,N\}$.


Next, in order to have an adaptive estimation of 
$\tilde{w}^o$ at each node $k$, we include the corresponding local aggregate estimate $\tilde{\phi}_{k}^{(i-1)}$ into the local LMS-type adaptive algorithm, at each node $k$.  
Therefore, the resulting diffusion-based strategy can be described as
\begin{gather}\label{eps:eq27}
\begin{split}
\left \{
\begin{array}{ll}
\textrm{Combination step:} \\
\tilde{\phi}_{k}^{(i-1)}=\sum_{j \in \mathcal{N}_{k,i-1}} \widetilde{C}_{k,j} \, \tilde{\psi}_{j}^{(i-1)}  \\
\\
\textrm{Adaptation step:} \\
\tilde{\psi}_k^{(i)} = \tilde{\phi}_{k}^{(i-1)}  - \mu_k \mathbf{\widetilde{U}}_{k,i}^H \left[ \mathbf{d}_{k,i} - \mathbf{\widetilde{U}}_{k,i} \, \tilde{\phi}_{k}^{(i-1)} \right ]  \\
\end{array}
\right .
\end{split}
\end{gather}
with $i \geq 1$, $\{\tilde{\psi}_{j}^{(0)}\}_{j \in \mathcal{N}_{k,0}}$ equal to some initial guesses, $\widetilde{C}_{k,j}$ defined in~\eqref{eps:eq13} and $\mu_k> 0$  is a suitably chosen positive step-size parameter.


Due to the structure of the augmented regressors $\mathbf{\widetilde{U}}_{k,i}$ defined in~\eqref{eps:eq6}, a careful analysis of~\eqref{eps:eq27} reveals that, only $2$ %
sub-vectors of $\tilde{\psi}_k^{(i)}$ are updated at each time instant $i$, when a specific node $k$ performs the adaptation step of~\eqref{eps:eq27}. In particular, according to~\eqref{eps:eq7} and~\eqref{eps:eq6}, only the sub-vectors associated with the local estimates of $w^o$ 
and $\xi_k^o$ at node $k$ and time $i$, denoted as $\psi_k^{(i)} = \psi_{k,w}^{(i)}$ 
and $\xi_k^{(i)}=\psi_{k,\xi_{k}}^{(i)}$, respectively, are updated based on the measurements $\{d_{k,i}, U_{k,i} \}$ and the corresponding aggregate estimates at time $i-1$, i.e., $\phi_{k,w}^{(i-1)}$ 
and $\phi_{k,\xi_{k}}^{(i-1)}$. 
The previous fact allows to set the subsequent equalities in the combination coefficients 
\begin{gather}\label{eps:eq28}
\begin{split}
c_{k,j}^{\xi_{m}} = 0 & \textrm{ if } k \neq j \textrm{ or }  k \neq m .
\end{split}
\end{gather}
These equalities together with~\eqref{eps:eq24} show that $c_{k,k}^{\xi_{k}} = 1$ for each node $k$. Hence, a node $k$ does not essentially cooperate with any other node when estimating its vector of local parameters $\xi_k^o$. This is due to the fact that no other node $j$ performs measurements where the vector $\xi_k$ is involved. 
Finally, we obtain the Combine-then-Adapt (CTA) diffusion-based LMS algorithm 
summarized below\\
\rule{\linewidth}{0.5mm} \\[-0.5mm]
\textbf{CTA Diffusion-based LMS for NSPE (CTA D-NSPE)}\\[-2mm]
\rule{\linewidth}{0.5mm}
\begin{itemize}
\item Start with some random guesses $\psi_{k}^{(0)}$ 
and $\xi_k^{(0)}$ at each node $k \in \{1,2,\ldots,N\}$ .
\item 
Choose a $N \times N$ combination matrix $C_{w}$ whose elements in each row $k$, i.e., $\{c_{k,j}^{w}\}_{j=1}^{N}$, satisfy~\eqref{eps:eq22}.
\item At each time $i$, for each $k \in \{1,2,\ldots,N\}$, execute
\item[] - Combination step:
\begin{gather}\label{eps:eq30a}
\begin{split}
\phi_{k,w}^{(i-1)} = \sum_{j \in \mathcal{N}_{k,i-1}} c_{k,j}^{w} \, \psi_{j}^{(i-1)}
\end{split}
\end{gather}
\item[] - Adaptation step:
\begin{gather}\label{eps:eq30}
\begin{split}
\begin{bmatrix}\psi_{k}^{(i)} \\ 
\xi_k^{(i)} \end{bmatrix}  
= \begin{bmatrix}\phi_{k,w}^{(i-1)} \\ 
 \xi_k^{(i-1)} \end{bmatrix}  + \mu_k \, U_{k,i}^H \left [ d_{k,i} - U_{k,i}  \begin{bmatrix}\phi_{k,w}^{(i-1)} \\ 
\xi_k^{(i-1)} \end{bmatrix}   \right]
\end{split}
\end{gather}
\end{itemize}
 \vspace{-0.25cm}
\rule{\linewidth}{0.5mm}\\[-2mm]


Now, let us consider that each node $k$ firstly performs the adaptation step 
and afterwards, 
it solves its local weighted least squares problem given in~\eqref{eps:eq15}. Then, by following a derivation that is analogous to the one undertaken for the CTA D-NSPE scheme and that has been omitted for the sake of brevity, we can obtain the subsequent Adapt-then-Combine (ATC) diffusion-based LMS algorithm. \\
\rule{\linewidth}{0.5mm} \\[-0.5mm]
\textbf{ATC Diffusion-based LMS for NSPE (ATC D-NSPE)}\\[-2mm]
\rule{\linewidth}{0.5mm}
\begin{itemize}
\item Start with some random guesses $\phi_{k,w}^{(0)}$ 
and $\xi_k^{(0)}$ at each node $k \in \{1,2,\ldots,N\}$ .
\item 
Choose a $N \times N$ combination matrix $C_{w}$ whose elements in each row $k$, i.e. $\{c_{k,j}^{w}\}_{j=1}^{N}$, satisfy~\eqref{eps:eq22}.
\item At each time $i$, for each $k \in \{1,2,\ldots,N\}$, execute
\item[] - Adaptation step:
\begin{gather}\label{eps:eq31}
\begin{split}
\begin{bmatrix}\psi_{k}^{(i)} \\ 
\xi_k^{(i)} \end{bmatrix}  
= \begin{bmatrix}\phi_{k,w}^{(i-1)} \\ 
\xi_k^{(i-1)} \end{bmatrix}  + \mu_k \, U_{k,i}^H \left [ d_{k,i} - U_{k,i}  \begin{bmatrix}\phi_{k,w}^{(i-1)} \\ 
\xi_k^{(i-1)} \end{bmatrix}   \right]
\end{split}
\end{gather}
\item[] - Combination step:
\begin{gather}\label{eps:eq31a}
\begin{split}
\phi_{k,w}^{(i)} = \sum_{j \in \mathcal{N}_{k,i}} c_{k,j}^{w} \, \psi_{j}^{(i)}
\end{split}
\end{gather}
\end{itemize}
 \vspace{-0.25cm}
\rule{\linewidth}{0.5mm}\\[-2mm]


Both diffusion-based NSPE algorithms are scalable in terms of computational burden and energy resources. On the one hand, regarding the computational complexity, at each time instant, each node $k$ only needs to update 
3 vectors whose dimensions are independent of the number of nodes. According to~\eqref{eps:eq30a}-\eqref{eps:eq31a}, these 
 vectors are $\phi_{k,w}^{(i-1)}$, $\psi_{k}^{(i)}$ 
and $\xi_k^{(i)}$, 
consequently, a total of $2 \, M_g 
+ M_l$ parameters are updated at node $k$ and any time instant $i$. On the other hand, at each time instant $i$, each node $k$ is required to transmit one 
vector, $\psi_{k}^{(i)}$, whose dimensions are again independent of the number of nodes.

\emph{Remark on convergence:} Let us comment briefly on the convergence in the mean of the two schemes. Assume that the combination matrix $C$ consists of $N \times N$ blocks, where each block $C(k,j)= \mathrm{diag}\{  c^{ w}_{k,j} I_{M_g},  c_{k,j}^{\xi_{k}} I_{M_l} \}$, and where the coefficients $c^{ w}_{k,j}$  $ c_{k,j}^{\xi_{k}}$ satisfy~\eqref{eps:eq22},~\eqref{eps:eq24} and~\eqref{eps:eq28}, i.e, $C 1=1$.
Furthermore, if every stepsize $\mu_k$ satisfies $0 < \mu_k < 2/\lambda_{\mathrm{max}} \{R_{U_{k,i}}\}$, where $\lambda_{\mathrm{max}}\{A\}$ equals the maximum eigenvalue of the positive definite matrix $A$,
then every local estimate at time instant $i$, i.e., $\text{col} \{\bold{\psi}_{k}^{(i)}, \bold{\xi}_k^{(i)}\}$, generated by  the CTA (or ATC) diffusion NSPE LMS algorithm converges in the mean to the optimal parameter $w^o_k$. 

\section{Simulations}
\label{sec:sec4}

At this point, we will compare the LMS-based ATC D-NSPE and CTA D-NSPE schemes with an LMS-based non-cooperative strategy, in a scenario of cooperative spectrum sensing in cognitive radio network (see~\cite{bogdanovic2013a}). We emphasize that we do not compare our schemes with the diffusion strategies designed for a scenario where $w_k^o=w^o$ for all $k \in \{1,2,\ldots,N\}$. Note that the comparison would not be fair since the latter strategies, e.g.,  \cite{cattivelli2010diffusion}, 
were not designed to estimate parameters of local interest.

Assume an environment shared by $Q$ primary users (PU) and $N$ secondary users (SU). In addition to PUs, for each SU we also assume a different local interference (LI) source. The goal for each SU is to estimate the aggregated spectrum transmitted by all the PUs and its own LI. Each node $k$, at time $i$, takes a set of measurements of the received power spectral density (PSD)  over $L$ frequency samples $\{f_m\}_{m=1}^L$, based on a basis expansion model that yields the following linear vector model~\cite{di2011bio},~\cite[Section 2.4]{sayed2012diffusion}:
\begin{gather}\label{eps:n3}
\begin{split}
\mathbf{d}_{k,i} &= \mathbf{U}_{k,i} \bar{w}_k^{o} + \mathbf{v}_{k,i} 
\end{split}
\end{gather}
where $\mathbf{d}_{k,i}$ is the measurement vector and $\mathbf{v}_{k,i}$ denotes noise with zero mean and covariance matrix $R_{v_{k},i}$ of dimension $L \times L$. 
The observation matrix relative to node $k$ is $\mathbf{U}_{k,i} = \left [ b_{k,i}^T(f_m) \right ]_{m=1}^L$, of dimensions $L \times (Q+1)J$ with $L > (Q+1)J$, whose rows are the evaluation of basis functions at frequency $f_m$. The overall coefficient vector $\bar{w}_k^{o} = \mathrm{col} \left \{ w_1^{o}, \ldots , w_Q^{o}, \xi_k^{o} \right \} \in \mathbb{R}^{(Q+1)J} $ collects the expansion coefficient vectors of all PUs  $\{ w^o_q\}^Q_{q=1}$ and one of LI, i.e., $\xi_k^o$, each of size $J$.

\begin{figure}[t]

\begin{minipage}[b]{\linewidth}
  \centering
  \centerline{\includegraphics[width=7cm]{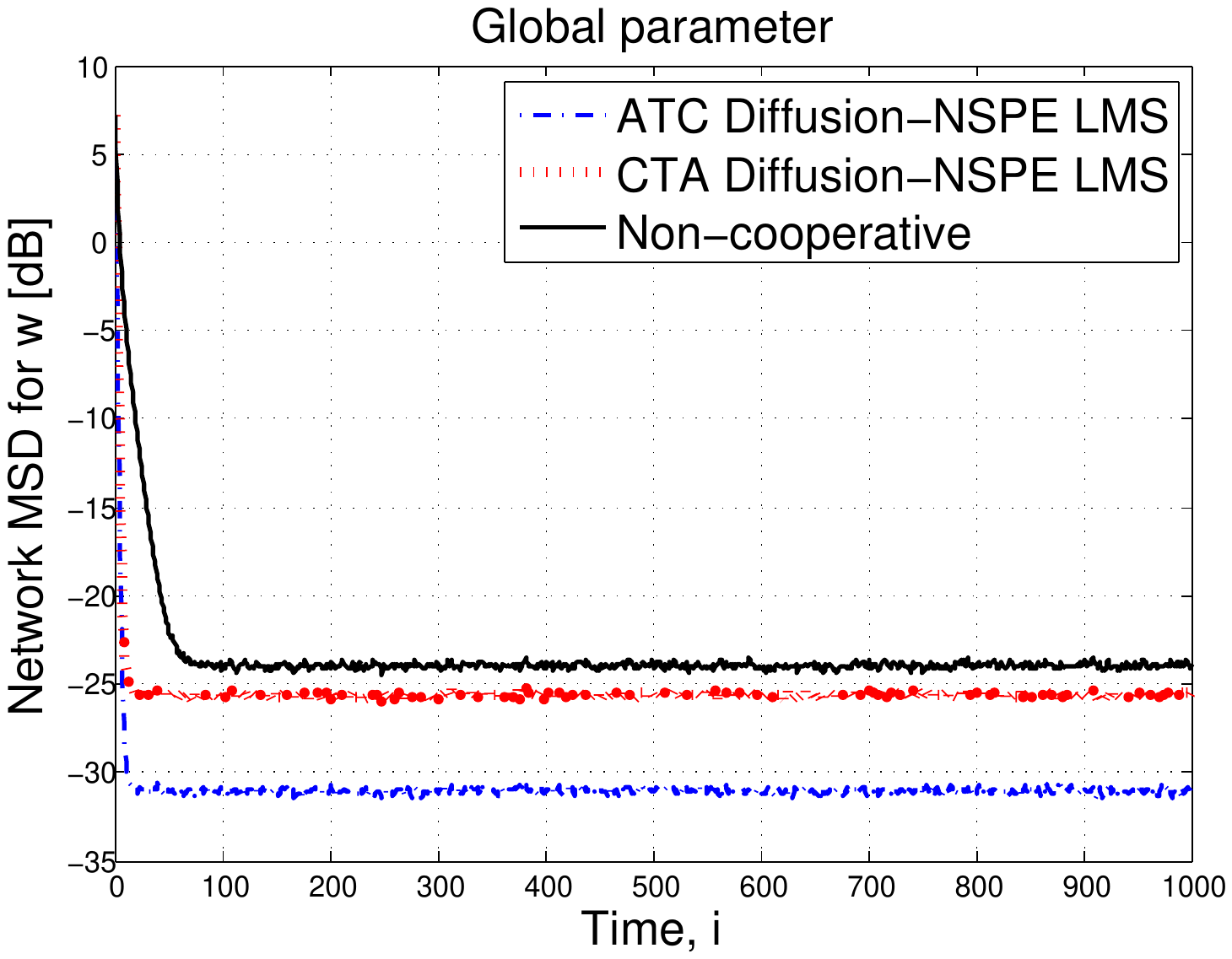}}
  \vspace{0.50cm}
\end{minipage}
%


\begin{minipage}[b]{\linewidth}
  \centering
  \centerline{\includegraphics[width=7.1cm]{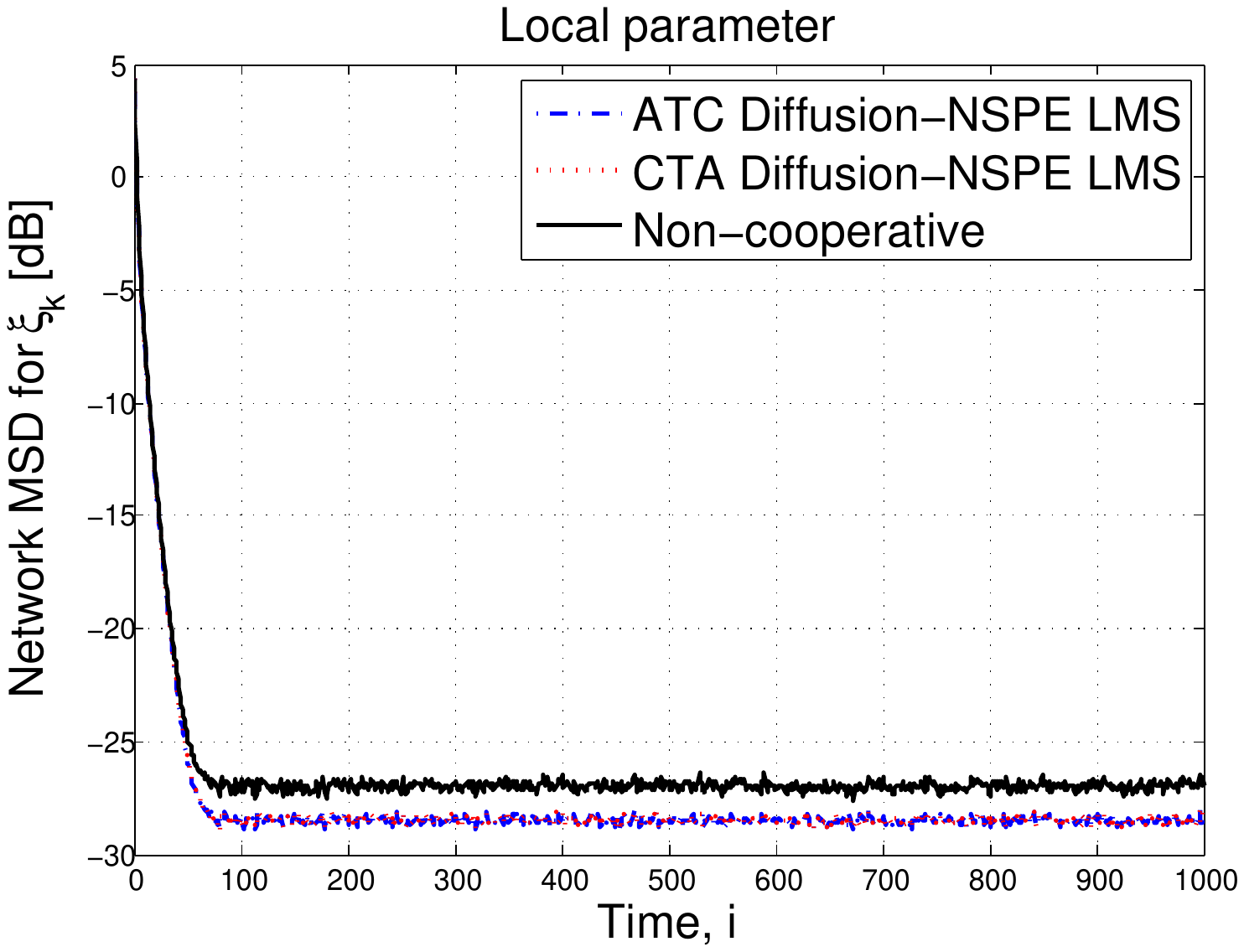}}
\end{minipage}
\hfill
%
\vspace{-0.5cm}
\caption{Learning behavior of network MSD.}
\label{fig:fig3}
 \vspace{-0.25cm}
\end{figure}

In Fig.~\ref{fig:fig3}, we plot the learning behavior of the three schemes in terms of the network mean-square deviation (MSD) associated with the estimation of $w^o$ 
 and $\xi_k^o$. Each network MSD is the result of averaging the local MSDs associated with the estimation of $w^o$ and $\xi_k^o$ at each node. 
To generate each plot, we have averaged the results over 50 independent experiments where we assumed $Q=2$ PUs, $N=10$ SUs and $J=16$ Gaussian basis functions. 
Furthermore, we have considered that each SU scans $L=80$ channels between 30 MHz and 45 MHz. The D-NSPE schemes use a simple averaging combination rule, where each SU cooperates with its 4 neighbors. The step-size is the same for all three schemes, i.e., $\mu= 3 \cdot 10^{-4}$ . 
Due to the cooperation between the nodes, we observe that the two proposed schemes outperform the non-cooperative one, especially when estimating $w^o$. 
Although there is no exchange of estimates of $\xi_k^o$ throughout the network, the D-NSPE schemes have enhanced performance in comparison with the non-cooperative strategy. This is a consequence of the coupling between the two estimation tasks undertaken by D-NSPE.

\section{Conclusions}
\label{sec:sec5}

We have presented a distributed adaptive algorithm that is suitable for solving node-specific parameter estimation problems in a network where the estimation interests of each node consist of a set of local parameters and a set of network global parameters. To do so, 
the proposed schemes employ local LMS algorithms allowing each node to estimate 
its set of local parameters. Coupled with all these local estimation processes, the estimation of the parameters of global interest is undertaken by a diffusion-based LMS that is implemented by all nodes of the network under a specific formulation, Combine-then-Adapt or Adapt-then-Combine. To conclude, computer simulations have been provided to show the effectiveness of the proposed schemes in the context of cooperative spectrum sensing in cognitive radio networks.
.

\bibliographystyle{IEEEbib}
\bibliography{references}

\begin{thebibliography}{10}

\bibitem{olfati2004}
R.~Olfati-Saber and R.~M. Murray,
\newblock ``Consensus problems in networks of agents with switching topology
  and time-delays,''
\newblock {\em IEEE Transactions on Automatic Control}, vol. 49, no. 9, pp.
  1520--1533, 2004.

\bibitem{bertsekasparallel}
D.~P. Bertsekas and J.~N. Tsitsiklis,
\newblock {\em {Parallel and distributed computation: numerical methods}},
\newblock Athena Scientific, Singapore, 1997.

\bibitem{schizas2009distributed}
I.~D. Schizas, G.~Mateos, and G.~B. Giannakis,
\newblock ``Distributed lms for consensus-based in-network adaptive
  processing,''
\newblock {\em IEEE Transactions on Signal Processing}, vol. 57, no. 6, pp.
  2365--2382, 2009.

\bibitem{dimakis2010gossip}
A.~G. Dimakis, S.~Kar, J.~M.~F. Moura, M.~G. Rabbat, and A.~Scaglione,
\newblock ``Gossip algorithms for distributed signal processing,''
\newblock {\em Proceedings of the IEEE}, vol. 98, no. 11, pp. 1847--1864, 2010.

\bibitem{lopes2007incremental}
C.~G. Lopes and A.~H. Sayed,
\newblock ``Incremental adaptive strategies over distributed networks,''
\newblock {\em IEEE Transactions on Signal Processing}, vol. 55, no. 8, pp.
  4064--4077, 2007.

\bibitem{li2010distributed}
L.~Li, J.~A. Chambers, C.~G. Lopes, and A.~H. Sayed,
\newblock ``Distributed estimation over an adaptive incremental network based
  on the affine projection algorithm,''
\newblock {\em IEEE Transactions on Signal Processing}, vol. 58, no. 1, pp.
  151--164, 2010.

\bibitem{cattivelli2010diffusion}
F.~S. Cattivelli and A.~H. Sayed,
\newblock ``Diffusion {L}{M}{S} strategies for distributed estimation,''
\newblock {\em IEEE Transactions on Signal Processing}, vol. 58, no. 3, pp.
  1035--1048, 2010.

\bibitem{chouvardas2011}
S.~Chouvardas, K.~Slavakis, and S.~Theodoridis,
\newblock ``Adaptive robust distributed learning in diffusion sensor
  networks,''
\newblock {\em IEEE Transactions on Signal Processing}, vol. 59, no. 10, pp.
  4692--4707, 2011.

\bibitem{sayed2013_magazine}
A.~H. Sayed, S-Y. Tu, J.~Chen, X.~Zhao, and Z.~J. Towfic,
\newblock ``Diffusion strategies for adaptation and learning over networks: an
  examination of distributed strategies and network behavior,''
\newblock {\em IEEE Signal Processing Magazine}, vol. 30, no. 3, pp. 155--171,
  2013.

\bibitem{bertrand2010distributed}
A.~Bertrand and M.~Moonen,
\newblock ``Distributed adaptive node-specific signal estimation in fully
  connected sensor networks - part {I}: Sequential node updating,''
\newblock {\em IEEE Transactions on Signal Processing}, vol. 58, no. 10, pp.
  5277--5291, 2010.

\bibitem{bertrand2010distributed_smlt}
A.~Bertrand and M.~Moonen,
\newblock ``Distributed adaptive node-specific signal estimation in fully
  connected sensor networks - part {II}: Simultaneous and asynchronous node
  updating,''
\newblock {\em IEEE Transactions on Signal Processing}, vol. 58, no. 10, pp.
  5292--5306, 2010.

\bibitem{kekatos2012distributed}
V.~Kekatos and G.~B. Giannakis,
\newblock ``Distributed robust power system state estimation,''
\newblock {\em Submitted to IEEE Transactions on Power Systems}, 2012 [Online].
  Available: http://arxiv.org/abs/1204.0991.

\bibitem{chen2012distributed}
J.~Chen and A.~H. Sayed,
\newblock ``{Distributed Pareto-optimal solutions via diffusion adaptation},''
\newblock in {\em IEEE Statistical Signal Processing Workshop, 2012. SSP
  2012.}, 2012, pp. 648--651.

\bibitem{zhao2012clustering}
X.~Zhao and A.~H. Sayed,
\newblock ``Clustering via diffusion adaptation over networks,''
\newblock in {\em 3rd International Workshop on Cognitive Information
  Processing, 2012. (CIP 2012)}, 2012, pp. 1--6.

\bibitem{abdolee2012diffusion}
R.~Abdolee, B.~Champagne, and A.~H. Sayed,
\newblock ``{Diffusion LMS for source and process estimation in sensor
  networks},''
\newblock in {\em IEEE/SP 17th Workshop on Statistical Signal Processing, 2012.
  SSP 2012}, 2012, pp. 165--168.

\bibitem{bogdanovic2013a}
N.~Bogdanovic, J.~Plata-Chaves, and K.~Berberidis,
\newblock ``{Distribtued incremental-based LMS for node-specific parameter
  estimation over adaptive networks},''
\newblock in {\em IEEE 38th International Conference on Acoustics, Speech and
  Signal Processing, 2013. ICASSP 2013}, 2013.

\bibitem{platachaves2013a}
J.~Plata-Chaves, N.~Bogdanovic, and K.~Berberidis,
\newblock ``{Distribtued incremental-based RLS for node-specific parameter
  estimation over adaptive networks},''
\newblock in {\em IEEE 21st European Signal Conference, 2013. EUSIPCO 2013},
  2013.

\bibitem{sayed2011adaptive}
A.~H. Sayed,
\newblock {\em Adaptive filters},
\newblock Wiley-IEEE Press, 2011.

\bibitem{lopes2008diffusion}
C.~G. Lopes and A.~H. Sayed,
\newblock ``Diffusion least-mean squares over adaptive networks: Formulation
  and performance analysis,''
\newblock {\em IEEE Transactions on Signal Processing}, vol. 56, no. 7, pp.
  3122--3136, 2008.

\bibitem{di2011bio}
P.~Di~Lorenzo, S.~Barbarossa, and A.~H. Sayed,
\newblock ``{Bio-inspired swarming for dynamic radio access based on diffusion
  adaptation},''
\newblock in {\em IEEE 19th European Signal Conference, 2011. EUSIPCO 2011},
  2012, pp. 1--6.

\bibitem{sayed2012diffusion}
A.~H. Sayed,
\newblock ``Diffusion adaptation over networks,''
\newblock {\em To appear in E-Reference Signal Processing, R. Chellapa and S.
  Theodoridis, Eds., Elsevier, 2013}, 2012 [Online]. Available:
  http://arxiv.org/abs/1205.4220.

\end{thebibliography}

\end{document}